\documentstyle[12pt]{article}
\advance\hoffset by -7mm
\makeatletter
\@addtoreset{equation}{section}
\makeatother
\renewcommand{\theequation}{\thesection.\arabic{equation}}
\renewcommand{\title}[1]{\null\vspace{25mm}

\noindent{\Large{\bf #1}}\vspace{10mm}

\noindent {\large By }}
\newcommand{\authors}[1]{\noindent{\large #1}\vspace{20mm}

}
\newcommand{\address}[1]{\noindent #1\vspace{5mm}

}
\renewcommand{\abstract}[1]{\vspace{17mm}

\noindent{\small{\em Abstract.} #1}\vspace{2mm}

}
\newcommand{\journal}[4]{{\em #1~}{\bf #2}\,(19#3)\,#4;}
\newcommand{\aihp}{\journal {Ann. Inst. Henri Poincar\'e}}
\newcommand{\hpa}{\journal {Helv. Phys. Acta}}

\newcommand{\pr}{\journal {Phys. Rev.}}

\newcommand{\jmp}{\journal {J. Math. Phys.}}

\newcommand{\cmp}{\journal {Comm. Math. Phys.}}
\newcommand{\cqg}{\journal {Class. Quantum Grav.}}

\newcommand{\np}{\journal {Nucl. Phys.}}
\newcommand{\pl}{\journal {Phys. Lett.}}

\newcommand{\prepsec}{\journal {Phys. Reports (Review Section
                                of Phys. Letters)}}

\newcommand{\nc}{\journal {Nuovo Cim.}}

\newcommand{\annp}{\journal {Ann. Phys. (N.Y.)}}

\makeatletter
\@addtoreset{equation}{section}
\makeatother
\renewcommand{\theequation}{\thesection.\arabic{equation}}

\newcommand{\lp}{\left(}\newcommand{\rp}{\right)}
\newcommand{\lc}{\left[}\newcommand{\rc}{\right]}
\newcommand{\lac}{\left\{}\newcommand{\rac}{\right\}}

\newcommand{\D}{\Delta}
\renewcommand{\a}{\alpha}

\renewcommand{\d}{\delta}

\newcommand{\x}{\xi}
\renewcommand{\l}{\lambda} 
\newcommand{\m}{\mu}

\renewcommand{\o}{\omega} \renewcommand{\O}{\Omega}

\newcommand{\r}{\rho}
\newcommand{\s}{\sigma}

\renewcommand{\AA}{{\cal A}}

\newcommand{\GG}{{\cal G}}

\newcommand{\LL}{{\cal L}}

\newcommand{\NN}{{\cal N}}

\newcommand{\QQ}{{\cal Q}}

\newcommand{\complex}{{\kern .1em {\raise .47ex
\hbox {$\scriptscriptstyle |$}}
    \kern -.4em {\rm C}}}
\newcommand{\real}{{{\rm I} \kern -.19em {\rm R}}}
\newcommand{\rational}{{\kern .1em {\raise .47ex
\hbox{$\scripscriptstyle |$}}
    \kern -.35em {\rm Q}}}
\renewcommand{\natural}{{\vrule height 1.6ex width
.05em depth 0ex \kern -.35em {\rm N}}}

\newcommand{\sla}{\raise.15ex\hbox{$/$}\kern -.57em}

\newcommand{
\twiddle}{\lower.9ex\rlap{$\kern -.1em\scriptstyle\sim$}}

\newcommand{\equ}[1]{(\ref{#1})}
\newcommand{\eq}{\begin{equation}}
\newcommand{\eqn}[1]{\label{#1}\end{equation}}
\newcommand{\eea}{\end{eqnarray}}
\newcommand{\eqa}{\begin{eqnarray}}
\newcommand{\eqan}[1]{\label{#1}\end{eqnarray}}
\newcommand{\ba}{\begin{array}}
\newcommand{\ea}{\end{array}}
\newcommand{\eqac}{\begin{equation}\begin{array}{rcl}}
\newcommand{\eqacn}[1]{\end{array}\label{#1}\end{equation}}

\def\non{\nonumber\\}

\def\6{\partial}
\def\ti{\tilde}
\def\wti{\widetilde}
\def\h{\eta}
\def\th{\theta}
\def\ph{\phi}
\def\={\!\!\!&=&\!\!\!}
\def\+{\!\!\!&&\!\!\!+~}
\def\-{\!\!\!&&\!\!\!-~}

\begin{document}

\setcounter{page}{0}
\thispagestyle{empty}\hspace*{\fill} REF. TUW 94-13

\title{Yang-Mills gauge anomalies in the presence of gravity with
       torsion}
\authors{O. Moritsch\footnote{Work supported in part by the
         ``Fonds zur F\"orderung der Wissenschaftlichen Forschung''
         under Contract Grant Number P9116-PHY.},
         M. Schweda and
         T. Sommer\footnote{Work supported in part by the
         ``Fonds zur F\"orderung der Wissenschaftlichen Forschung''
         under Contract Grant Number P10268-PHY.}}
\address{Institut f\"ur Theoretische Physik,
         Technische Universit\"at Wien\\
         Wiedner Hauptstra\ss e 8-10, A-1040 Wien (Austria)}
\abstract{
The BRST transformations for the Yang-Mills gauge fields in the
presence of gravity with torsion are discussed by using the
so-called Maurer-Cartan horizontality conditions.
With the help of an operator $\d$ which allows to decompose the
exterior spacetime derivative as a BRST commutator we solve the
Wess-Zumino consistency condition corresponding to invariant
Chern-Simons terms and gauge anomalies.
}


\newpage
\section{Introduction}

Gauge fields play an important role in many theories of physical
interest, especially in the unification of all fundamental
interactions.
Anomalies can occur when the symmetries of the classical theory
are not preserved at the quantum level.
It is well-known that the search of the invariant Lagrangians and
of the anomalies corresponding to a given set of field
transformations can be done in a purely algebraic way by solving
the BRST consistency equation, a generalization of the famous
Wess-Zumino consistency condition~\cite{zum}, in the space of the
integrated local field polynomials.

\noindent
This leads to study the non-trivial solutions of the following
cohomology problem
\eq
s\D=0~~~,~~~\D \not= s\hat{\D} \ ,
\eqn{CE}
where $\D$ and $\hat{\D}$ are integrated
local field polynomials and $s$ is the nilpotent BRST operator.
Setting $\D=\int\AA$, condition \equ{CE} translates into the local
equation
\eq
s\AA+d\QQ=0 \ ,
\eqn{LE}
where $\QQ$ is some local polynomial
and $d=dx^{\mu}\6_{\mu}$ denotes the
exterior spacetime derivative which,
together with the BRST operator $s$, obeys:
\eq
s^{2}=d^{2}=sd+ds=0 \ .
\eqn{NIL}
$\AA$ is said non-trivial if
\eq
\AA \not= s\hat\AA+d\hat\QQ \ ,
\eqn{NONTRIVIAL}
with $\hat\AA$ and $\hat\QQ$ local polynomials.
In this case the integral of $\AA$ on spacetime, $\int\AA$,
identifies a cohomology class of the BRST operator $s$ and,
according to its ghost number,
it corresponds to an invariant Lagrangian
(ghost number zero) or to an anomaly (ghost number one).

The local
equation \equ{LE}, due to the relations \equ{NIL} and to the
algebraic Poincar\'e Lemma~\cite{cotta,dragon},
is easily seen to generate a tower of descent equations
\eqa
&&s\QQ+d\QQ^{1}=0\non
&&s\QQ^{1}+d\QQ^{2}=0\non
&&~~~~~.....\non
&&~~~~~.....\non
&&s\QQ^{k-1}+d\QQ^{k}=0\non
&&s\QQ^{k}=0 \ ,
\eqan{LADDER}
with $\QQ^{i}$ local field polynomials.

As it has been well-known for several years, these equations can be
solved by using a transgression procedure based on the so-called
{\it Russian formula}
\cite{witten,baulieu,dvio,tmieg,ban,ginsparg,tonin1,stora,brandt,fb}.
More recently an alternative way of finding non-trivial solutions of
the ladder \equ{LADDER} has been proposed by S.P. Sorella and
successfully applied to the study of the Yang-Mills gauge
anomalies~\cite{silvio}.
The method is based on the introduction of an operator $\d$ which
allows to express the exterior derivative $d$ as a BRST commutator,
i.e.:
\eq
d=-[s,\d] \ .
\eqn{DECOMP}
One easily verifies that, once the decomposition \equ{DECOMP} has
been found, successive applications of the operator $\d$ on the
polynomial $\QQ^{k}$ which solves the last equation of the tower
\equ{LADDER} give an explicit non-trivial solution for the higher
cocycles $\QQ^{k-1}, ....., \QQ^{1},
\QQ,$ and $\AA$.

Remark however, that in the case of pure Yang-Mills gauge theory
the algebra between the operators $d$,$s$ and $\d$ is not
closed~\cite{silvio}.
In order to close the algebra, one has to introduce a further
operator $\GG$ defined by the following commutator relation:
\eq
2\GG = [d,\d] \ .
\eqn{DEF_G_OPERATOR}
In our case of a Yang-Mills gauge theory in curved spacetime
with torsion this $\GG$-operator does not appear,
because of the explicit presence of diffeomorphisms.

The decomposition \equ{DECOMP} represents one of the most
interesting features of the topological
field theories~\cite{schwarz,birmingham} and of the bosonic string
and superstring in the Beltrami and Super-Beltrami
parametrization~\cite{manfred}.
Actually, one has also found that the operator $\d$ corresponds to
a vector supersymmetry in Witten's topological Yang-Mills field
theory~\cite{mom}.
A further remarkable fact is that solving the last equation of the
tower \equ{LADDER} is only a problem of local BRST cohomology
instead of a modulo-$d$ one.
One sees then that, due to the operator $\d$, the study of the
cohomology of $s$ modulo $d$ is essentially
reduced to the study of the local cohomology of $s$ which, in turn,
can be systematically analyzed by using the powerful technique of
the spectral sequences~\cite{dixon}. Recently, as
proven in~\cite{tataru}, the solutions
obtained by utilizing the decomposition \equ{DECOMP} turn out to be
completely equivalent to that based on the {\it Russian formula},
i.e. they differ only by trivial cocycles (see also~\cite{fb}).

The aim of this work is twofold. In a first step it will be
demonstrated that the decomposition \equ{DECOMP} can be successfully
extended to Yang-Mills gauge theory in a curved spacetime,
and that it holds also in the presence of torsion.
In the second step, we will see that the operator $\d$ gives an
elegant and straightforward way of classifying the cohomology
classes of the full BRST operator in any spacetime dimension.
As shown in~\cite{oms,diss}, the eq.\equ{DECOMP} allows a
cohomological interpretation of the cosmological constant, of
Lagrangians for pure Einstein gravity and generalizations including
also torsion, as well as gravitational Chern-Simons terms and
anomalies.
The incorporation of scalar matter fields and Weyl transformations
has been done in~\cite{diss}, where one can also find a discussion
of Weyl anomalies. Here we want to concentrate us on Chern-Simons
terms and gauge anomalies.

The further analysis is based on the geometrical formalism
introduced by L. Baulieu and J. Thierry-Mieg~\cite{baulieu,tmieg}
which allows to reinterpret the BRST transformations
as Maurer-Cartan horizontality conditions.
In particular, this formalism turns out to be very useful in the
case of gravity~\cite{baulieu,tmieg}, since it naturally includes
the torsion.
In addition, it allows to formulate the diffeomorphism
transformations as local translations in the tangent space
by means of the introduction
of the ghost field $\h^{a}=\x^{\mu}e_{\mu}^{a}$ where $\x^{\mu}$
denotes the usual diffeomorphism ghost and $e_{\mu}^{a}$ is the
vielbein\footnote{As usual, Latin and Greek indices refer to the
tangent space and to the Euclidean spacetime.}.

We recall also that the BRST formulation of gravity with torsion
has already been proposed by~\cite{okubo,lee} in order to
study the quantum aspects of gravity.
In particular, the authors of~\cite{lee} discussed a four
dimensional torsion Lagrangian, with $GL(4,R)$ as the gauge group,
which is able to reproduce the Einstein gravity in the low
energy limit. These BRST transformations could be taken as the
starting point for a purely cohomological algebraic analysis
without any reference to a particular Lagrangian.
Furthermore, our choice of adopting the Maurer-Cartan formalism is
due to the fact that when combined with the introduction of the
translation ghost $\h^{a}$ it will give
us the possibility of a fully tangent space formulation of gravity.

This step, as we shall see in details, will allow to
introduce the decomposition \equ{DECOMP} in a very simple way
and will produce an elegant and compact formula (see Section 3)
for expressing the whole solution of the BRST descent equations,
our aim being that of giving a cohomological interpretation of the
Chern-Simons terms and of the gauge anomalies in any spacetime
dimension.
Moreover, the explicit presence of the torsion $T$ and of the
translation ghost $\h^{a}$ gives the possibility of
introducing an algebraic BRST setup which turns out to be
different from that obtained from the analysis of
Brandt et al.~\cite{brandt},
where similar techniques have been used.

Finally, we stress that we always refer to the elementary fields,
i.e. the vielbein $e$, the Lorentz connection $\o$, the
Yang-Mills gauge field $A$, the Riemann tensor $R$,
the torsion $T$, and the Yang-Mills gauge
field strength $F$ as unquantized classical fields,
as done in~\cite{werneck},
which when coupled to some matter fields (scalars or fermions)
give rise to an effective action whose quantum expansion reduces to
the one-loop order.

The paper is organized as follows. In Section 2 we will introduce
the so-called Maurer-Cartan horizontality conditions for Yang-Mills
gauge fields in the presence of gravity with torsion.
In particular, the BRST transformations for local Lorentz rotations,
diffeomorphisms, and gauge transformations are derived in a complete
tangent space formalism.
In Section 3 the operator $\d$ is introduced and we show how it can
be used to solve the descent equations \equ{LADDER}. Furthermore,
one can find a discussion about the geometrical meaning
of the decomposition \equ{DECOMP}.
Section 4 is devoted to the study of the Chern-Simons terms and
gauge anomalies in any dimension. In Section 5 we briefly discuss
the appearance of the $\GG$-operator \equ{DEF_G_OPERATOR}
and some detailed calculations are given in the final Appendices.

\section{Maurer-Cartan horizontality conditions}

The aim of this section is to derive the set of BRST transformations
for the Yang-Mills gauge fields in the presence of gravity with
torsion from Maurer-Cartan horizontality
conditions~\cite{baulieu,tmieg}.
In a first step this geometrical formalism is used to discuss
the simpler case of non-abelian Yang-Mills theory~\cite{baulieu2}.


\subsection{Pure Yang-Mills gauge field theory}

The BRST transformations of the 1-form gauge connection
$A^{A}=A^{A}_{\mu}dx^{\mu}$ and the 0-form
ghost field $c^{A}$ are given by
\footnote{Here, capital Latin indices are denoting gauge indices.}
\eqa
sA^{A}\=dc^{A}+f^{ABC}c^{B}A^{C} \ , \non
sc^{A}\=\frac{1}{2}f^{ABC}c^{B}c^{C} \ ,
\eqan{BRST_YM}
with
\eq
s^{2}=0 \ ,
\eqn{NILPOTENCY_YM}
where $f^{ABC}$ are the structure constants of the corresponding
gauge group ${\bf G}$.
As usual, the adopted grading is given by the sum of the
form degree and of the ghost number. In this sense, the fields
$A^{A}$ and $c^{A}$ are both of degree one,
their ghost number being respectively zero and one.
A $p$-form with ghost number $q$ will be denoted by
$\O^{q}_{p}$, its total grading being $(p+q)$.
The 2-form field strength $F^{A}$ is given by
\eq
F^{A}=\frac{1}{2}F^{A}_{\mu\nu}dx^{\mu}dx^{\nu}
=dA^{A}+\frac{1}{2}f^{ABC}A^{B}A^{C} \ ,
\eqn{FIELDSTRENGTH_YM}
and
\eq
dF^{A}=f^{ABC}F^{B}A^{C} \ ,
\eqn{BIANCHI_YM}
is its Bianchi identity.
In order to reinterpret the BRST transformations \equ{BRST_YM} as a
Maurer-Cartan horizontality condition we introduce the combined
gauge-ghost field
\eq
\wti{A}^{A}=A^{A}+c^{A} \ ,
\eqn{A_TILDE_YM}
and the generalized nilpotent differential operator
\eq
\wti{d}=d-s \ , ~~~\wti{d}^{2}=0 \ .
\eqn{d_TILDE_YM}
Notice that both $\wti{A}^{A}$ and $\wti{d}$ have degree one.
The nilpotency of $\wti{d}$ in \equ{d_TILDE_YM} just implies the
nilpotency of $s$ and $d$,
and furthermore fulfills the anticommutator
relation
\eq
\lac s,d \rac = 0 \ .
\eqn{ANTICOMMUTATOR_YM}

Let us introduce also the degree-two field strength $\wti{F}^{A}$:
\eq
\wti{F}^{A}=\wti{d}\wti{A}^{A}
+\frac{1}{2}f^{ABC}\wti{A}^{B}\wti{A}^{C} \ ,
\eqn{F_TILDE_YM}
which, from eq.\equ{d_TILDE_YM}, obeys the generalized Bianchi
identity
\eq
\wti{d}\wti{F}^{A}=f^{ABC}\wti{F}^{B}\wti{A}^{C} \ .
\eqn{GENERAL_BIANCHI_YM}
The Maurer-Cartan horizontality condition
reads then
\eq
\wti{F}^{A}=F^{A} \ .
\eqn{MCHC_YM}
Now it is very easy to check that the BRST transformations
\equ{BRST_YM} can be obtained from the horizontality condition
\equ{MCHC_YM} by simply expanding $\wti{F}^{A}$ in terms of the
elementary fields $A^{A}$ and $c^{A}$ and collecting the terms
with the same form degree and ghost number.


\subsection{Yang-Mills gauge fields in the presence of gravity}

In order to generalize the horizontality condition \equ{MCHC_YM}
to the case of Yang-Mills gauge fields in the presence of gravity
let us first specify the functional space
the BRST operator $s$ acts upon. The latter is chosen to be the
space of local polynomials which depend on the 1-forms
$(e^{a},\o^{a}_{~b},A^{A})$, where $e^{a}$, $\o^{a}_{~b}$, and
$A^{A}$ being respectively the vielbein, the Lorentz connection,
and the Yang-Mills gauge field
\eqa
e^{a} \= e^{a}_{\mu}dx^{\mu} \ , \non
\o^{a}_{~b} \= \o^{a}_{~b\mu}dx^{\mu} \ , \non
A^{A} \= A^{A}_{\mu}dx^{\mu} \ ,
\eqan{OF}
and on the $2$-forms $(T^{a},R^{a}_{~b},F^{A})$,
whereby $T^{a}$, $R^{a}_{~b}$, and $F^{A}$
denoting the torsion, the Riemann tensor, and the Yang-Mills
field strength
\eqa
T^{a} \= \frac{1}{2}T^{a}_{\mu\nu}dx^{\mu}dx^{\nu}
=de^{a}+\o^{a}_{~b}e^{b}
=De^{a} \ , \non
R^{a}_{~b} \= \frac{1}{2}R^{a}_{~b\mu\nu}dx^{\mu}dx^{\nu}
=d\o^{a}_{~b}+
\o^{a}_{~c}\o^{c}_{~b} \ , \non
F^{A} \= \frac{1}{2}F^{A}_{\mu\nu}dx^{\mu}dx^{\nu}
=dA^{A}+\frac{1}{2}f^{ABC}A^{B}A^{C} \ ,
\eqan{TF}
with the covariant derivative
\eq
D=d+\o+A \ .
\eqn{COVD}
The tangent space indices $(a,b,c,...)$ in eqs.\equ{OF} and
\equ{TF} are referred to the group $SO(N)$, $N$ being the
dimension of the Euclidean spacetime.

Applying the exterior derivative $d$ to both sides of eq.\equ{TF}
one gets the Bianchi identities
\eqa
DT^{a} \= dT^{a}+\o^{a}_{~b}T^{b}=R^{a}_{~b}e^{b} \ , \non
DR^{a}_{~b} \= dR^{a}_{~b}+\o^{a}_{~c}R^{c}_{~b}
-\o^{c}_{~b}R^{a}_{~c}=0 \ , \non
DF^{A} \=dF^{A}+f^{ABC}A^{B}F^{C}=0 \ .
\eqan{BI}

To write down the gravitational Maurer-Cartan horizontality
conditions for this case one introduces a further ghost,
as done in~\cite{baulieu,tmieg},
the local translation ghost $\h^{a}$ having ghost number one
and a tangent space index.
As explained in~\cite{tmieg}, the field $\h^{a}$ represents the
ghost of local translations in the tangent space. See also the
discussion of~\cite{mielke} based on an affine approach to gravity.

The local translation ghost $\h^{a}$ can be related~\cite{tmieg} to
the ghost of local diffeomorphism $\x^{\mu}$ by the relation
\eq
\x^{\mu} = E^{\mu}_{a} \h^{a}~~~,~~~
\h^{a} = \x^{\mu}e^{a}_{\mu} \ ,
\eqn{ETA}
where $E^{\mu}_{a}$ denotes the inverse of the vielbein
$e^{a}_{\mu}$, i.e.
\eqa
e^{a}_{\mu}E^{\mu}_{b} \= \d^{a}_{b} \ ,\non
e^{a}_{\mu}E^{\nu}_{a} \= \d^{\nu}_{\mu} \ .
\eqan{VIEL_ORTHO}
Proceeding now as for the pure Yang-Mills case, one defines the
nilpotent differential operator $\wti{d}$ of degree one:
\eq
\wti{d}=d-s \ ,
\eqn{EXTD}
and the generalized vielbein-ghost field $\ti{e}^{a}$, the extended
Lorentz connection $\wti{\o}^{a}_{~b}$, and the generalized
non-abelean Yang-Mills gauge field $\wti{A}$
\eqa
\ti{e}^{a} \= e^{a}+\h^{a} \ , \non
\wti{\o}^{a}_{~b} \= \widehat{\o}^{a}_{~b}+\th^{a}_{~b} \ , \non
\wti{A}^{A} \= \widehat{A}^{A}+c^{A} \ ,
\eqan{EVIEL}
where $\widehat{\o}^{a}_{~b}$
and $\widehat{A}^{A}$ are given by
\eqa
\widehat{\o}^{a}_{~b}\=\o^{a}_{~bm}\ti{e}^{m}
=\o^{a}_{~b}+\o^{a}_{~bm}\h^{m} \ , \non
\widehat{A}^{A}\=A^{A}_{m}\ti{e}^{m}
=A^{A}+A^{A}_{m}\h^{m} \ ,
\eqan{HCO}
with the 0-forms $\o^{a}_{~bm}$\footnote{Remark that the
0-form $\o^{a}_{~bm}$ does not possess any symmetric or
antisymmetric property with respect to the lower indices $(bm)$.}
and $A^{A}_{m}$ defined by the expansion of the 0-form
connection $\o^{a}_{~b\mu}$ and the 0-form Yang-Mills gauge field
$A^{A}_{\m}$ in terms of the vielbein $e^{a}_{\mu}$,
i.e.:
\eqa
\o^{a}_{~b\mu}\=\o^{a}_{~bm}e^{m}_{\mu} \ , \non
A^{A}_{\mu}\=A^{A}_{m}e^{m}_{\mu} \ .
\eqan{SPINC}
As it is well-known, the last formulas stem from the fact that the
vielbein formalism allows to transform locally the spacetime
indices of an arbitrary
tensor $\NN_{\mu\nu\r\s...}$ into flat tangent space indices
$\NN_{abcd...}$ by means of the expansion
\eq
\NN_{\mu\nu\r\s...}=\NN_{abcd...}
e^{a}_{\mu}e^{b}_{\nu}e^{c}_{\r}e^{d}_{\s}...  \ .
\eqn{WORLD}
Vice versa one has
\eq
\NN_{abcd...}=\NN_{\mu\nu\r\s...}
E^{\mu}_{a}E^{\nu}_{b}E^{\r}_{c}E^{\s}_{d}...  \ .
\eqn{TANG}
According to eqs.\equ{TF}, the generalized torsion field,
the generalized Riemann tensor, and the generalized Yang-Mills
field strength are given by
\eqa
\wti{T}^{a} \= \wti{d}\ti{e}^{a}+\wti{\o}^{a}_{~b}\ti{e}^{b}
=\wti{D}\ti{e}^{a} \ , \non
\wti{R}^{a}_{~b} \= \wti{d}\wti{\o}^{a}_{~b}+\wti{\o}^{a}_{~c}
\wti{\o}^{c}_{~b} \ , \non
\wti{F}^{A}\=\wti{d}\wti{A}^{A}
+\frac{1}{2}f^{ABC}\wti{A}^{B}\wti{A}^{C} \ ,
\eqan{DEFTWO}
and are easily seen to obey the generalized Bianchi identities
\eqa
\wti{D}\wti{T}^{a} \= \wti{d}\wti{T}^{a}
+\wti{\o}^{a}_{~b}\wti{T}^{b}
=\wti{R}^{a}_{~b}\ti{e}^{b} \ , \non
\wti{D}\wti{R}^{a}_{~b} \= \wti{d}\wti{R}^{a}_{~b}
+\wti{\o}^{a}_{~c}\wti{R}^{c}_{~b}
-\wti{\o}^{c}_{~b}\wti{R}^{a}_{~c}=0 \ , \non
\wti{D}\wti{F}^{A}\=\wti{d}\wti{F}^{A}
+f^{ABC}\wti{A}^{B}\wti{F}^{C}=0 \ ,
\eqan{GBI}
with
\eq
\wti{D}=\wti{d}+\wti{\o}+\wti{A}
\eqn{GENERAL_COVD}
the generalized covariant derivative.

With these definitions the Maurer-Cartan horizontality conditions
for the Yang-Mills gauge fields in the presence of gravity (with
non-vanishing torsion) may be expressed in the following way:
{\it $\ti{e}$ and all its generalized covariant
exterior differentials can be expanded over $\ti{e}$ with
classical coefficients},
\eqa
\label{MCG1}
\ti{e}^{a}\=\d^{a}_{b}\ti{e}^{b} \equiv horizontal \ , \\
\label{MCG2}
\wti{T}^{a}(\ti{e},\wti{\o})
\=\frac{1}{2}T^{a}_{mn}(e,\o)\ti{e}^{m}\ti{e}^{n}
\equiv horizontal \ , \\
\label{MCG3}
\wti{R}^{a}_{~b}(\wti{\o})
\=\frac{1}{2}R^{a}_{~bmn}(\o)\ti{e}^{m}\ti{e}^{n}
\equiv horizontal \ , \\
\label{MCG4}
\wti{F}^{A}(\wti{A})
\=\frac{1}{2}F^{A}_{mn}(A)\ti{e}^{m}\ti{e}^{n}
\equiv horizontal \ .
\eqan{MAURER_CARTAN_HC}
Through eq.\equ{WORLD}, the 0-forms $T^{a}_{mn}$,
$R^{a}_{~bmn}$, and $F^{A}_{mn}$
are defined by the vielbein expansion of the 2-forms of the
torsion, the Riemann tensor, and the Yang-Mills field strength
of eq.\equ{TF},
\eqa
T^{a}\=\frac{1}{2}T^{a}_{mn}e^{m}e^{n} \ , \non
R^{a}_{~b}\=\frac{1}{2}R^{a}_{~bmn}e^{m}e^{n} \ , \non
F^{A}\=\frac{1}{2}F^{A}_{mn}e^{m}e^{n} \ ,
\eqan{TWOFORM}
and the 0-form $D_{m}$ of the covariant exterior derivative $D$
is given by
\eq
D=e^{m}D_{m} \ .
\eqn{ED}
Notice also that eqs.\equ{HCO} are nothing but the horizontality
conditions for the Lorentz connection and the Yang-Mills gauge field
expressing the fact that
$\widehat{\o}$ and $\widehat{A}$
themselfs can be expanded over $\ti{e}$.

Eqs.\equ{MCG1}-\equ{MCG4} define the Maurer-Cartan horizontality
conditions for the Yang-Mills gauge fields in the presence of
gravity and, when expanded in terms of
the elementary fields
$(e^{a}, \o^{a}_{~b}, A^{A}, \h^{a}, \th^{a}_{~b}, c^{A})$, give the
nilpotent BRST transformations corresponding to the diffeomorphism
transformations, the local Lorentz rotations, and
the gauge transformations.

For a better understanding of this point let us discuss in details
the horizontality condition \equ{MCG2} for the torsion.
Making use of eqs.\equ{EXTD}, \equ{EVIEL}, \equ{HCO} and of the
definition \equ{DEFTWO}, one verifies that eq.\equ{MCG2} gives
\eqa
&&de^{a}-se^{a}+d\h^{a}-s\h^{a}+\o^{a}_{~b}e^{b}
+\th^{a}_{~b}e^{b}\non
&&+~\o^{a}_{~b}\h^{b}+\th^{a}_{~b}\h^{b}
+\o^{a}_{~bm}\h^{m}e^{b}+\o^{a}_{~bm}\h^{m}\h^{b}\non
&&=\frac{1}{2}T^{a}_{mn}e^{m}e^{n}
+T^{a}_{mn}e^{m}\h^{n}+\frac{1}{2}T^{a}_{mn}\h^{m}\h^{n} \ ,
\eqan{EXPANSION}
from which, collecting the terms with the same form degree and
ghost number, one easily obtains the BRST transformations for the
vielbein $e^{a}$ and for the ghost $\h^{a}$:
\eqa
se^{a}\=d\h^{a}+\o^{a}_{~b}\h^{b}+\th^{a}_{~b}e^{b}
+\o^{a}_{~bm}\h^{m}e^{b}-T^{a}_{mn}e^{m}\h^{n} \ ,\non
s\h^{a}\=\th^{a}_{~b}\h^{b}+\o^{a}_{~bm}\h^{m}\h^{b}
-\frac{1}{2}T^{a}_{mn}\h^{m}\h^{n} \ .
\eqan{BRSE}
These equations, when rewritten in terms of the variable $\x^{\mu}$
of eq.\equ{ETA}, take the more familiar form
\eqa
se^{a}_{\mu}\=\th^{a}_{~b}e^{b}_{\mu}+\LL_{\x}e^{a}_{\mu} \ ,\non
s\x^{\mu}\=-\x^{\l}\6_{\l}\x^{\mu},
\eqan{BRS_WORLD}
where $\LL_{\x}$ denotes the ordinary Lie derivative along the
direction $\x^{\mu}$, i.e.
\eq
\LL_{\x}e^{a}_{\mu}=-\x^{\l}\6_{\l}e^{a}_{\mu}
-(\6_{\mu}\x^{\l})e^{a}_{\l} \ .
\eqn{LIE_DERIV}
It is apparent now that eq.\equ{BRSE} represents the tangent space
formulation of the usual BRST transformations corresponding to local
Lorentz rotations and diffeomorphisms.

One sees then that the Maurer-Cartan horizontality conditions
\equ{MCG1}-\equ{MCG4} together with eq.\equ{DEFTWO} carry in a very
simple and compact way all the information relative to the
gravitational Yang-Mills gauge algebra. It is easy indeed to expand
eqs.\equ{MCG1}-\equ{MCG4} in terms of $e^{a}$ and $\h^{a}$ and work
out the BRST transformations of the remaining fields
$(\o^{a}_{~b},A^{A},T^{a},R^{a}_{~b},F^{A},\th^{a}_{~b},c^{A})$.

However, in view of the fact that we will use as fundamental
variables
the 0-forms $(\o^{a}_{~bm}, A^{A}_{m}, T^{a}_{mn},
R^{a}_{~bmn}, F^{A}_{mn})$ rather than
the 1-forms $\o^{a}_{~b}$ and $A^{A}$ and the 2-forms
$T^{a}$, $R^{a}_{~b}$, and $F^{A}$ let us proceed by introducing the
partial derivative $\6_{a}$ with indices in the tangent space.
According to the formulas \equ{WORLD} and \equ{TANG}, the latter is
defined by
\eq
\6_{a} \equiv E^{\mu}_{a}\6_{\mu} \ ,
\eqn{TANGENT_DERIV}
and
\eq
\6_{\mu} = e^{a}_{\mu}\6_{a} \ ,
\eqn{WORLD_DERIV}
so that the intrinsic exterior differential $d$ becomes
\eq
d=dx^{\mu}\6_{\mu}=e^{a}\6_{a} \ .
\eqn{EXTER_DERIV}
Let us emphasize that the introduction of the operator $\6_{a}$ and
the use of the 0-forms $(\o^{a}_{~bm}, A^{A}_{m}, T^{a}_{mn},
R^{a}_{~bmn}, F^{A}_{mn})$
allows for a complete tangent space formulation of the gravitational
Yang-Mills gauge algebra.
This step, as we shall see later, turns out to be very
useful in the analysis of the corresponding BRST cohomology.
Moreover, as one can easily understand, the knowledge of the BRST
transformations
of the 0-form sector $(\o^{a}_{~bm}, A^{A}_{m}, T^{a}_{mn},
R^{a}_{~bmn}, F^{A}_{mn})$
together with the expansions \equ{SPINC}, \equ{TWOFORM} and the
equation \equ{BRSE} completely characterize the transformation
law of the forms $(\o^{a}_{~b}, A^{A}, T^{a},
R^{a}_{~b}, F^{A})$.

We remark however that, contrary to the case of the usual
spacetime derivative $\6_{\mu}$, the operator $\6_{a}$ does not
commute with the BRST operator
due to the explicit presence of the vielbein $e^{a}$.
One has:
\eq
[s,\6_{m}]=(\6_{m}\h^{k}-\th^{k}_{~m}-T^{k}_{mn}\h^{n}
-\o^{k}_{~mn}\h^{n}+\o^{k}_{~nm}\h^{n})\6_{k} \ ,
\eqn{COMM_1}
and
\eq
[\6_{m},\6_{n}]=-(T^{k}_{mn}+\o^{k}_{~mn}-\o^{k}_{~nm})\6_{k} \ .
\eqn{COMM_2}
Nevertheless, taking into account the vielbein transformation
\equ{BRSE}, one consistently verifies that
\eq
\{s,d\}=0 \ ,~~~d^{2}=0 \ .
\eqn{ANTICOMM_S_D}

\subsection{BRST transformations and Bianchi identities}

Let us finish this chapter by giving, for the convenience of the
reader, the BRST transformations and the Bianchi identities which
one can find by using the Maurer-Cartan horizontality conditions
\equ{MCG1}-\equ{MCG4} and from eqs.\equ{DEFTWO}, \equ{GBI} for
each form sector and ghost number.

\begin{itemize}

\item {\bf Form sector 2 $(T^{a}, R^{a}_{~b}, F^{A})$}

\eqa
sT^{a}\=\th^{a}_{~b}T^{b}+\o^{a}_{~bk}\h^{k}T^{b}
-R^{a}_{~b}\h^{b}\non
\+\o^{a}_{~b}T^{b}_{mn}e^{m}\h^{n}-R^{a}_{~bmn}e^{b}e^{m}\h^{n}
+(dT^{a}_{mn})e^{m}\h^{n}\non
\-T^{a}_{mn}e^{m}d\h^{n}+T^{a}_{mn}T^{m}\h^{n}
-T^{a}_{kn}\o^{k}_{~m}e^{m}\h^{n} \ , \non
sR^{a}_{~b}\=\th^{a}_{~c}R^{c}_{~b}-\th^{c}_{~b}R^{a}_{~c}
+\o^{a}_{~ck}\h^{k}R^{c}_{~b}-\o^{c}_{~bk}\h^{k}R^{a}_{~c}\non
\+\o^{a}_{~c}R^{c}_{~bmn}e^{m}\h^{n}
-\o^{c}_{~b}R^{a}_{~cmn}e^{m}\h^{n}+(dR^{a}_{~bmn})e^{m}\h^{n}\non
\+R^{a}_{~bmn}T^{m}\h^{n}
-R^{a}_{~bkn}\o^{k}_{~m}e^{m}\h^{n}
-R^{a}_{~bmn}e^{m}d\h^{n} \ , \non
sF^{A}\=(dF^{A}_{mn})e^{m}\h^{n}+F^{A}_{mn}T^{m}\h^{n}
-F^{A}_{mn}\o^{m}_{~k}e^{k}\h^{n}-F^{A}_{mn}e^{m}d\h^{n}\non
\+f^{ABC}c^{B}F^{C}+f^{ABC}A^{B}_{m}\h^{m}F^{C}
+f^{ABC}A^{B}F^{C}_{mn}e^{m}\h^{n} \ .
\eqan{FORMTWO}
For the Bianchi identities one has
\eqa
&&dT^{a}+\o^{a}_{~b}T^{b}=R^{a}_{~b}e^{b} \ , \non
&&dR^{a}_{~b}+\o^{a}_{~c}R^{c}_{~b}
-\o^{c}_{~b}R^{a}_{~c}=0 \ , \non
&&dF^{A}+f^{ABC}A^{B}F^{C}=0 \ .
\eqan{BIG}

\item {\bf Form sector 1 $(e^{a}, \o^{a}_{~b}, A^{A})$}

\eqa
se^{a}\=d\h^{a}+\o^{a}_{~b}\h^{b}+\th^{a}_{~b}e^{b}
+\o^{a}_{~bm}\h^{m}e^{b}
-T^{a}_{mn}e^{m}\h^{n} \ , \non
s\o^{a}_{~b}\=d\th^{a}_{~b}+\th^{a}_{~c}\o^{c}_{~b}
+\o^{a}_{~c}\th^{c}_{~b}
+(d\o^{a}_{~bm})\h^{m}+\o^{a}_{~bm}d\h^{m}\non
\+\o^{a}_{~c}\o^{c}_{~bm}\h^{m}+\o^{a}_{~cm}\h^{m}\o^{c}_{~b}
-R^{a}_{~bmn}e^{m}\h^{n} \ , \non
sA^{A}\=dc^{A}+(dA^{A}_{m})\h^{m}+A^{A}_{m}d\h^{m}
+f^{ABC}A^{B}c^{C}\non
\+f^{ABC}A^{B}A^{C}_{m}\h^{m}
-F^{A}_{mn}e^{m}\h^{n} \ .
\eqan{FORM1}


\item {\bf Form sector 0, ghost number 0
$(\o^{a}_{~bm}, A^{A}_{m}, T^{a}_{mn}, R^{a}_{~bmn}, F^{A}_{mn})$ }

\eqa
s\o^{a}_{~bm}\=-\6_{m}\th^{a}_{~b}+\th^{a}_{~c}\o^{c}_{~bm}
-\th^{c}_{~b}\o^{a}_{~cm}
-\th^{k}_{~m}\o^{a}_{~bk}-\h^{k}\6_{k}\o^{a}_{~bm} \ , \non
sA^{A}_{m}\=-\6_{m}c^{A}-f^{ABC}A^{B}_{m}c^{C}
-\th^{k}_{~m}A^{A}_{k}-\h^{k}\6_{k}A^{A}_{m} \ , \non
sT^{a}_{mn}\=\th^{a}_{~k}T^{k}_{mn}-\th^{k}_{~m}T^{a}_{kn}
-\th^{k}_{~n}T^{a}_{mk}-\h^{k}\6_{k}T^{a}_{mn} \ , \non
sR^{a}_{~bmn}\=\th^{a}_{~c}R^{c}_{~bmn}-\th^{c}_{~b}R^{a}_{~cmn}
-\th^{k}_{~m}R^{a}_{~bkn}-\th^{k}_{~n}R^{a}_{~bmk}
-\h^{k}\6_{k}R^{a}_{~bmn} \ , \non
sF^{A}_{mn}\=f^{ABC}c^{B}F^{C}_{mn}-\th^{k}_{~m}F^{A}_{kn}
-\th^{k}_{~n}F^{A}_{mk}-\h^{k}\6_{k}F^{A}_{mn} \ .
\eqan{0_FORMS}
The Bianchi identities \equ{BIG} are projected on the 0-form
torsion $T^{a}_{mn}$, on the 0-form curvature $R^{a}_{~bmn}$,
and on the 0-form Yang-Mills field strength $F^{A}_{mn}$
to give
\eqa
dT^{a}_{mn}\=(\6_{l}T^{a}_{mn})e^{l}\non
\=(R^{a}_{~lmn}+R^{a}_{~mnl}+R^{a}_{~nlm}\non
\-\o^{a}_{~bl}T^{b}_{mn}-\o^{a}_{~bm}T^{b}_{nl}
-\o^{a}_{~bn}T^{b}_{lm}\non
\+T^{a}_{kn}T^{k}_{ml}+T^{a}_{km}T^{k}_{ln}
+T^{a}_{kl}T^{k}_{nm}\non
\-T^{a}_{kn}\o^{k}_{~lm}-T^{a}_{km}\o^{k}_{~nl}
-T^{a}_{kl}\o^{k}_{~mn}\non
\+T^{a}_{kn}\o^{k}_{~ml}+T^{a}_{kl}\o^{k}_{~nm}
+T^{a}_{km}\o^{k}_{~ln}\non
\-\6_{m}T^{a}_{nl}-\6_{n}T^{a}_{lm})e^{l} \ , \non
dR^{a}_{~bmn}\=(\6_{l}R^{a}_{~bmn})e^{l}\non
\=(-\o^{a}_{~cl}R^{c}_{~bmn}-\o^{a}_{~cm}R^{c}_{~bnl}
-\o^{a}_{~cn}R^{c}_{~blm}\non
\+\o^{c}_{~bl}R^{a}_{~cmn}+\o^{c}_{~bm}R^{a}_{~cnl}
+\o^{c}_{~bn}R^{a}_{~clm}\non
\+R^{a}_{~bkn}T^{k}_{ml}+R^{a}_{~bkm}T^{k}_{ln}
+R^{a}_{~bkl}T^{k}_{nm}\non
\-R^{a}_{~bkn}\o^{k}_{~lm}-R^{a}_{~bkm}\o^{k}_{~nl}
-R^{a}_{~bkl}\o^{k}_{~mn}\non
\+R^{a}_{~bkn}\o^{k}_{~ml}+R^{a}_{~bkl}\o^{k}_{~nm}
+R^{a}_{~bkm}\o^{k}_{~ln}\non
\-\6_{m}R^{a}_{~bnl}-\6_{n}R^{a}_{~blm})e^{l} \ , \non
dF^{A}_{mn}\=(\6_{l}F^{A}_{mn})e^{l}\non
\=(f^{ABC}F^{B}_{mn}A^{C}_{l}+f^{ABC}F^{B}_{nl}A^{C}_{m}
+f^{ABC}F^{B}_{lm}A^{C}_{n}\non
\-F^{A}_{kn}T^{k}_{lm}-F^{A}_{kl}T^{k}_{mn}
-F^{A}_{km}T^{k}_{nl}\non
\+F^{A}_{kn}\o^{k}_{~ml}+F^{A}_{km}\o^{k}_{~ln}
+F^{A}_{kl}\o^{k}_{~nm}\non
\-F^{A}_{kn}\o^{k}_{~lm}-F^{A}_{kl}\o^{k}_{~mn}
-F^{A}_{km}\o^{k}_{~nl}\non
\-\6_{m}F^{A}_{nl}-\6_{n}F^{A}_{lm})e^{l} \ .
\eqan{0_BI}
One has also the equations
\eqa
d\o^{a}_{~bm}\=(\6_{n}\o^{a}_{~bm})e^{n}\non
\=(-R^{a}_{~bmn}+\o^{a}_{~cm}\o^{c}_{~bn}
-\o^{a}_{~cn}\o^{c}_{~bm}\non
\+\o^{a}_{~bk}T^{k}_{mn}-\o^{a}_{~bk}\o^{k}_{~nm}
+\o^{a}_{~bk}\o^{k}_{~mn}+\6_{m}\o^{a}_{~bn})e^{n} \ , \non
dA^{A}_{m}\=(\6_{n}A^{A}_{m})e^{n}\non
\=(-F^{A}_{mn}+f^{ABC}A^{B}_{m}A^{C}_{n}+A^{A}_{k}T^{k}_{mn}\non
\-A^{A}_{k}\o^{k}_{~nm}+A^{A}_{k}\o^{k}_{~mn}
+\6_{m}A^{A}_{n})e^{n} \ .
\eqan{0_DEF}

\item {\bf Form sector 0, ghost number 1
$(\h^{a}, \th^{a}_{~b}, c^{A})$ }

\eqa
s\h^{a}\=\o^{a}_{~bm}\h^{m}\h^{b}+\th^{a}_{~b}\h^{b}
-\frac{1}{2}T^{a}_{mn}\h^{m}\h^{n} \ , \non
s\th^{a}_{~b}\=\th^{a}_{~c}\th^{c}_{~b}
-\h^{k}\6_{k}\th^{a}_{~b} \ , \non
sc^{A}\=\frac{1}{2}f^{ABC}c^{B}c^{C}-\h^{k}\6_{k}c^{A} \ .
\eqan{FORMZERO}

\item {\bf Algebra between $s$ and $d$ }

{}From the above transformations it follows:
\eq
s^{2}=0 \ ,~~~d^{2}=0 \ ,
\eqn{S_D_ALGEBRA_1}
and
\eq
\{s,d\}=0 \ .
\eqn{S_D_ALGEBRA_2}

\end{itemize}

Let us conclude this section by making a remark about the use of
the variable $\h^a$. Observe that, when expressed in terms
of $\h^a$, the BRST transformation of the vielbein $e^a$
in \equ{FORM1} starts with a term linear
in the fields (i.e. the term $d\h^a$). This feature makes
the analogy between gravitational and gauge theories
more transparent.
Moreover, it suggests that one
may compute the local cohomology of the gravitational
BRST operator without expanding the vielbein $e^a$ around a
background, as shown in~\cite{dragon,fb}.

\section{Decomposition of the exterior derivative}

In this section we introduce the decomposition \equ{DECOMP} and
we show how it can be used to solve the ladder \equ{LADDER}.
To this purpose we introduce the operator $\d$ defined as
\eqa
\d\h^{a}\=-e^{a} \ , \non
\d\ph\=0~~~{\rm for}~~~\ph=(e, \o, A, T, R, F, \th, c) \ .
\eqan{DECETA}
It is easy to verify that $\d$ is of degree zero and that, together
with the BRST operator $s$, it obeys the following algebraic
relations:
\eq
[s,\d]=-d \ ,
\eqn{DEC}
and
\eq
[d,\d]=0 \ .
\eqn{EXCOM}
One sees from eq.\equ{DEC} that the operator $\d$ allows to
decompose the exterior derivative $d$ as a BRST commutator.
This property, as shown in~\cite{dragon}, gives an elegant and
simple procedure for solving the equations \equ{LADDER}.

Let us consider indeed the tower of descent equations which
originates from a local field polynomials $\O^{G}_{N}$
in the variables
($e^{a}$, $\o^{a}_{~bm}$, $A^{A}_{m}$, $T^{a}_{mn}$,
$R^{a}_{~bmn}$, $F^{A}_{mn}$, $\h^{a}$, $\th^{a}_{~b}$, $c^{A}$)
and their derivatives with ghost number $G$ and form degree $N$,
$N$ being the dimension of the spacetime,
\eqa
&&s\O^{G}_{N}+d\O^{G+1}_{N-1}=0\non
&&s\O^{G+1}_{N-1}+d\O^{G+2}_{N-2}=0\non
&&~~~~~.....\non
&&~~~~~.....\non
&&s\O^{G+N-1}_{1}+d\O^{G+N}_{0}=0\non
&&s\O^{G+N}_{0}=0 \ ,
\eqan{TOWER}
with $(\O^{G+1}_{N-1},...., \O^{G+N-1}_{1}, \O^{G+N}_{0})$ local
polynomials which, without loss of generality, will be always
considered as irreducible elements, i.e. they cannot be expressed as
the product of several factorized terms.
In particular, the ghost numbers $G=(0,1)$ correspond respectively
to an invariant Lagrangian and to an anomaly.

Thanks to the operator $\d$ and to the algebraic relations
\equ{DEC}-\equ{EXCOM}, in order to find a solution of the
ladder \equ{TOWER} it is sufficient to solve only the last equation
for the 0-form $\O^{G+N}_{0}$.
It is easy to check that, once a non-trivial solution for
$\O^{G+N}_{0}$
is known, the higher cocycles $\O^{G+N-q}_{q}$, $(q=1,...,N)$ are
obtained by repeated applications of the operator $\d$ on
$\O^{G+N}_{0}$, i.e.
\eq
\O^{G+N-q}_{q}=\frac{\d^{q}}{q!}\O^{G+N}_{0}~~~,~~~q=1,...,N~~~,
{}~~~G=(0,1) \ .
\eqn{HICO}
Let us emphasize also that solving the last equation of the tower
\equ{TOWER} is a problem of {\it local} BRST cohomology instead of a
modulo-$d$ one. One sees then, by means of the decomposition
\equ{DEC}, that the study of the cohomology of $s$ modulo $d$ is
reduced to the study of the local cohomology of $s$.
It is well-known indeed that, once a particular solution of the
descent equations \equ{TOWER} has been obtained, i.e. eq.\equ{HICO},
the search of the most general solution becomes essentially a
problem of local BRST cohomology.

Having discussed the role of the operator $\d$ in finding explicit
solutions of the descent equations \equ{TOWER}, let us turn now to
the study of its geometrical meaning.
As we shall see, this operator turns out to possess a quite simple
geometrical interpretation which will reveal an unexpected and so
far unnoticed elementary structure of the ladder \equ{TOWER}.

Let us begin by observing that all the cocycles $\O^{G+N-p}_{p}$
$(p=0,...,N)$ entering the descent equations \equ{TOWER} are of the
same degree (i.e. $(G+N)$), the latter being given by the sum of
the ghost number and of the form degree.

We can collect then, following~\cite{tataru},
all the $\O^{G+N-p}_{p}$
into a unique cocycle $\widehat{\O}$ of degree $(G+N)$ defined as
\eq
\widehat{\O}=\sum_{p=0}^{N}\O^{G+N-p}_{p} \ .
\eqn{COC_SUM}
This expression, using eq.\equ{HICO}, becomes
\eq
\widehat{\O}=\sum_{p=0}^{N}\frac{\d^{p}}{p!}\O^{G+N}_{0} \ ,
\eqn{SUM}
where the cocycle $\O^{G+N}_{0}$, according to its form degree,
depends only on the set of 0-form variables
$(\h^{a}, \th^{a}_{~b}, c^{A}, \o^{a}_{~bm}, A^{A}_{m}, T^{a}_{mn},
R^{a}_{~bmn}, F^{A}_{mn})$
and their tangent space derivatives $\6_{m}$.
Taking into account that under the action of the operator $\d$ the
form degree and the ghost number are respectively raised and lowered
by one unit and that in a spacetime of dimension $N$ a $(N+1)$-form
identically vanishes, it follows that eq.\equ{SUM} can be rewritten
in a more suggestive way as
\eq
\widehat{\O}=e^{\d}\O^{G+N}_{0}(\h^{a}, \th^{a}_{~b}, c^{A},
\o^{a}_{~bm}, A^{A}_{m}, T^{a}_{mn}, R^{a}_{~bmn}, F^{A}_{mn}) \ .
\eqn{EXPONENT}
Let us make now the following elementary but important remark.
As one can see from eq.\equ{DECETA}, the operator $\d$ acts as a
translation on the ghost $\h^{a}$ with an
amount given by $(-e^{a})$.
Therefor $e^{\d}$ has the simple effect of shifting $\h^{a}$ into
$(\h^{a}-e^{a})$. This implies that the cocycle \equ{EXPONENT} takes
the form
\eq
\widehat{\O}=\O^{G+N}_{0}(\h^{a}-e^{a}, \th^{a}_{~b}, c^{A},
\o^{a}_{~bm}, A^{A}_{m}, T^{a}_{mn}, R^{a}_{~bmn}, F^{A}_{mn}) \ .
\eqn{SHIFT}
This formula collects in a very elegant and simple expression the
solution of the descent equations \equ{TOWER}.

In particular, it states the important result that:

\begin{quote}
 {\it To find a non-trivial solution of the ladder} \equ{TOWER}
{\it it
is sufficient to replace the variable $\h^{a}$ with $(\h^{a}-e^{a})$
in the 0-form cocycle $\O^{G+N}_{0}$ which belongs to the local
cohomology of the BRST operator $s$.

The expansion
of $\O^{G+N}_{0}(\h^{a}-e^{a}, \th^{a}_{~b}, c^{A}, \o^{a}_{~bm},
A^{A}_{m}, T^{a}_{mn}, R^{a}_{~bmn}, F^{A}_{mn})$ in powers of the
1-form vielbein $e^{a}$ yields then all the searched
cocycles $\O^{G+N-p}_{p}$.}
\end{quote}

Let us conclude by remarking that expression
\equ{SHIFT} represents a deeper understanding of the algebraic
properties of the gravitational ladder \equ{TOWER} and of the role
played by the vielbein $e^{a}$ and the associated ghost $\h^{a}$.

\section{Chern-Simons terms and gauge anomalies}

In this section we will focus on the invariants and anomalies
in the Yang-Mills sector. Other examples of non-trivial explicit
solutions of the consistency condition \equ{CE} in the
presence of gravity can be found e.g. in~\cite{oms,diss}.

For the sake of clarity and to make contact with the results
obtained in~\cite{silvio,fb},
let us discuss in detail the construction of the
three-dimensional Chern-Simons term.
In this case the tower \equ{TOWER} takes
the form
\eqa
&&s\O^{0}_{3}+d\O^{1}_{2}=0 \ ,\non
&&s\O^{1}_{2}+d\O^{2}_{1}=0 \ ,\non
&&s\O^{2}_{1}+d\O^{3}_{0}=0 \ ,\non
&&s\O^{3}_{0}=0 \ ,
\eqan{TOWER_CS_3}
where, according to eq.\equ{HICO},
\eqa
&&\O^{2}_{1}=\d\O^{3}_{0} \ ,\non
&&\O^{1}_{2}=\frac{\d^{2}}{2!}\O^{3}_{0} \ ,\non
&&\O^{0}_{3}=\frac{\d^{3}}{3!}\O^{3}_{0} \ .
\eqan{HICO_3}
In order to find a solution for $\O^{3}_{0}$ we use the
following redefined ghost variables:
\eqa
\hat{c}^{A}\=A^{A}_{m}\h^{m}+c^{A} \ , \non
\hat{F}^{A}\=\frac{1}{2}F^{A}_{mn}\h^{m}\h^{n} \ ,
\eqan{C_HAT}
which, from eq.\equ{DECETA}, transform as
\eq
\d\hat{c}^{A}=-A^{A}~~~,~~~\d\hat{F}^{A}=-F^{A}_{mn}e^{m}\h^{n} \ .
\eqn{DELTA_C_HAT}
For the cocycle $\O^{3}_{0}$ one then has
\eq
\O^{3}_{0}=\frac{1}{3!}f^{ABC}\hat{c}^{A}\hat{c}^{B}\hat{c}^{C}
-\hat{F}^{A}\hat{c}^{A} \ ,
\eqn{ZEROFORM_CS_3}
from which $\O^{2}_{1}$, $\O^{1}_{2}$, and $\O^{0}_{3}$ are
computed to be
\eqa
\O^{2}_{1}=-\frac{1}{2}f^{ABC}A^{A}\hat{c}^{B}\hat{c}^{C}
+\hat{F}^{A}A^{A}+F^{A}_{mn}e^{m}\h^{n}\hat{c}^{A} \ ,
\eqan{COCYCLES_CS_3}
\eqa
\O^{1}_{2}=\frac{1}{2}f^{ABC}A^{A}A^{B}\hat{c}^{C}
-F^{A}_{mn}e^{m}\h^{n}A^{A}-F^{A}\hat{c}^{A} \ ,
\eqan{CHERNTWO}
\eqa
\O^{0}_{3}=-\frac{1}{6}f^{ABC}A^{A}A^{B}A^{C}+F^{A}A^{A} \ .
\eqan{CHERNTHREE}
In particular, expression \equ{CHERNTHREE} gives the familiar
three-dimensional Chern-Simons term.
Finally, let us remark that the
cocycle $\O^{1}_{2}$ of eq.\equ{CHERNTWO},
when referred to dimension $N=2$, reduces to the expression
\eq
\O^{1}_{2}=-(dA^{A})c^{A}~~~,~~~for~N=2 \ ,
\eqn{GAUGE_ANOMALY_2}
which directly gives the two-dimensional gauge anomaly.

\noindent
Analogous, for the 0-form cocycle $\O^{5}_{0}$ in five dimensions
one gets
\eqa
\O^{5}_{0}\=d^{ABC}\hat{c}^{A}\hat{F}^{B}\hat{F}^{C}
-\frac{1}{4}d^{ABC}f^{CKL}
\hat{c}^{A}\hat{F}^{B}\hat{c}^{K}\hat{c}^{L} \non
\+\frac{1}{40}d^{ABC}f^{BMN}f^{CKL}\hat{c}^{A}\hat{c}^{M}
\hat{c}^{N}\hat{c}^{K}\hat{c}^{L} \ ,
\eqan{ZEROFORM_CS_5}
which leads to the five-dimensional Chern-Simons term
\eqa
\O^{0}_{5}\=-d^{ABC}A^{A}F^{B}F^{C}
+\frac{1}{4}d^{ABC}f^{CKL}A^{A}F^{B}A^{K}A^{L} \non
\-\frac{1}{40}d^{ABC}f^{BMN}f^{CKL}A^{A}A^{M}A^{N}A^{K}A^{L} \ ,
\eqan{CHERNFIVE}
where $d^{ABC}$ is the invariant totally symmetric tensor of
rank three
\eq
d^{ABC}=\frac{1}{2}Tr\lp T^{A}\lac T^{B}, T^{C} \rac\rp \ ,
\eqn{D_TENSOR}
and $T^{A}$ are the generators of a finite
representation of the gauge group ${\bf G}$.

The corresponding cocycle $\O^{1}_{4}$ is given by
\eqa
\O^{1}_{4}\=-d^{ABC}\hat{c}^{A}F^{B}F^{C}
-2d^{ABC}A^{A}F^{B}F^{C}_{mn}e^{m}\h^{n} \non
\+\frac{1}{2}d^{ABC}f^{CMN}A^{A}F^{B}A^{M}\hat{c}^{N}
+\frac{1}{4}d^{ABC}f^{CMN}A^{A}F^{B}_{mn}e^{m}\h^{n}A^{M}A^{N} \non
\+\frac{1}{4}d^{ABC}f^{CMN}\hat{c}^{A}F^{B}A^{M}A^{N}
-\frac{1}{8}d^{ABC}f^{BMN}f^{CKL}\hat{c}^{A}A^{M}A^{N}A^{K}A^{L} \ ,
\eqan{OMEGA_1_4}
which, when referred to dimension $N=4$, reduces to the expression
\eqa
\O^{1}_{4}\=-d^{ABC}c^{A}F^{B}F^{C}
+\frac{1}{2}d^{ABC}f^{CMN}A^{A}F^{B}A^{M}c^{N} \non
\+\frac{1}{4}d^{ABC}f^{CMN}c^{A}F^{B}A^{M}A^{N}
-\frac{1}{8}d^{ABC}f^{BMN}f^{CKL}c^{A}A^{M}A^{N}A^{K}A^{L} \ ,
\eqan{OMEGA_1_4_YM}
which represents the four-dimensional gauge anomaly.

Generally, in $(2k-1)$ dimensions one has for the 0-form
cocycle $\O^{2k-1}_{0}$ (see Appendix B) the formula in
matrix notation, i.e. $\hat{c}=\hat{c}^{A}T^{A}$ and
$\hat{F}=\hat{F}^{A}T^{A}$,
\eqa
\O^{2k-1}_{0}\=-k~Tr\int\limits_{0}^{1} dt~\hat{c}
\hat{F}(t)^{k-1} \ ,
\eqan{CS_FORMULA_ZERO}
where $\hat{F}^{A}(t)$ is defined as
\eqa
\hat{F}^{A}(t)\=t\hat{F}^{A}
+\frac{1}{2}t(t-1)f^{ABC}\hat{c}^{B}\hat{c}^{C} \ .
\eqan{F_HAT_T}
According to eq.\equ{HICO} one easily finds the corresponding
$(2k-1)$-dimensional Chern-Simons terms:
\eqa
\O^{0}_{2k-1}\=\frac{\d^{2k-1}}{(2k-1)!}\O^{2k-1}_{0} \non
\=k~Tr\int\limits_{0}^{1} dt~AF(t)^{k-1} \ ,
\eqan{CS_GENERAL}
where $F^{A}(t)$ is given by
\eqa
F^{A}(t)\=tF^{A}
+\frac{1}{2}t(t-1)f^{ABC}A^{B}A^{C} \ .
\eqan{F_(T)}

Let us make here a remark concerning the cancelation of the terms
which contain the local translation ghost $\h$ in the cocycles
$\O^{1}_{2k-2}$ for the anomalies in $(2k-2)$ dimensions.
The gauge anomalies have their cohomological origin in the 0-form
cocycles $\O^{2k-1}_{0}$ of \equ{CS_FORMULA_ZERO}, but one has to
note that the dimension is now reduced to $(2k-2)$.
Taking into account that in a $N$-dimensional spacetime the product
of $(N+1)$ ghost fields $\h^{a}$ automatically vanishes, it is
easily seen that $\O^{2k-1}_{0}$ can only contain at most $(2k-2)$
local translation ghosts $\h$. Therefore, from the decomposition
\equ{DECETA} and the solution of the ladder \equ{HICO}
follow that the gauge anomalies $\O^{1}_{2k-2}$ cannot contain
the local translation ghost $\h$. So, all terms with $\h$
in $\O^{1}_{2k-2}$ are identically zero and one can simply
drop them.

\section{The $\GG$-operator}

This section is devoted to the analysis of the influence of
diffeomorphisms at solving the descent equations \equ{TOWER}.
In the presence of diffeomorphisms one has with \equ{EXCOM}
\eq
2\GG = [d,\d] = 0 \ .
\eqn{DEF_GG}
{}From the decomposition \equ{DEC} follows for the redefined
gauge ghost \equ{C_HAT}:
\eq
\d\hat{c}^{A}=-A^{A} \ .
\eqn{DEC_C_HAT}
By making the limit $\h\rightarrow 0$ one obtains a new
decomposition
\eq
\lim_{\h\rightarrow 0}\d\hat{c}^{A}=\bar{\d} c^{A}=-A^{A} \ .
\eqn{DEC_C}
We remark, that this decomposition implies now a non-vanishing
$\bar{\GG}$-operator. To see this explicitly let us discuss
in detail the decomposition of the 1-form gauge field $A^{A}$
for the cases with and without diffeomorphisms. In the first case,
with the help of eq.\equ{FORM1}, one has
\eqa
-\lc s,\d \rc A^{A}\=\d sA^{A}
=\d(d\hat{c}^{A}+f^{ABC}\hat{c}^{B}A^{C}
-F^{A}_{mn}e^{m}\h^{n}) \non
\=\d d\hat{c}^{A}+2dA^{A}=dA^{A} \ ,
\eqan{DEC_A_ETA}
and one gets
\eq
\d d\hat{c}^{A}=-dA^{A}=d\d\hat{c}^{A}
{}~~~\Longleftrightarrow~~~\GG \hat{c}^{A}=0 \ .
\eqn{DELTA_DC_HAT}
In the absence of diffeomorphisms, i.e. in the limit
$\h\rightarrow 0$,
above result holds no more:
\eqa
-\lc s,\bar{\d} \rc A^{A}\=\bar{\d} sA^{A}
=\bar{\d}(dc^{A}+f^{ABC}c^{B}A^{C}) \non
\=\bar{\d} dc^{A}-f^{ABC}A^{B}A^{C}=dA^{A} \ ,
\eqan{DEC_A}
which gives
\eq
\bar{\d} dc^{A}=dA^{A}+f^{ABC}A^{B}A^{C}\not= d\bar{\d} c^{A}
{}~~~\Longrightarrow~~~\bar{\GG} c^{A}\not= 0 \ .
\eqn{DELTA_DC}
So one can see that the diffeomorphisms carry in some sense the
action of the $\bar{\GG}$-operator.

\section{Conclusion}

The algebraic structure of Yang-Mills gauge field theory in the
presence of gravity with torsion has been analyzed
in the context of the Maurer-Cartan horizontality formalism by
introducing an operator $\delta$ which allows
to decompose the exterior spacetime derivative as a BRST commutator.
Such a decomposition gives a simple and elegant way of solving
the Wess-Zumino consistency condition corresponding to invariant
Lagrangians and anomalies. The same technique can be
applied to the characterization of the Weyl anomalies~\cite{next}.

\section*{Acknowledgements}

We would like to thank Fran\c{c}ois Gieres,
Mauricio W. de Oliveira and Silvio P. Sorella
for all the useful discussions and comments.

\section*{Appendices}

\section*{A~~~Chern-Simons formula}

\setcounter{equation}{0}
\renewcommand{\theequation}{A.\arabic{equation}}

To find a general formula for Chern-Simons terms
in $(2k-1)$ dimensions one introduces
the interpolating 1-form gauge field
$A^{A}(t)$, $t\in[0,1]$ in the following way~\cite{stora}:
\eq
A^{A}(t)=tA^{A} \ ,
\eqn{A_T}
with $A^{A}(0)=0$ and $A^{A}(1)=A^{A}$.
The associated 2-form field strength $F^{A}(t)$ is then given by
\eqa
F^{A}(t)\=dA^{A}(t)+\frac{1}{2}f^{ABC}A^{B}(t)A^{C}(t)  \non
\=tdA^{A}+\frac{1}{2}t^{2}f^{ABC}A^{B}A^{C} \non
\=tF^{A}+\frac{1}{2}t(t-1)f^{ABC}A^{B}A^{C} \ ,
\eqan{F_T}
with $F^{A}(0)=0$ and $F^{A}(1)=F^{A}$.
For simplicity one can define a covariant derivative $D_{t}$
with respect to $A(t)$ according to
\eq
D_{t}=d+A(t) \ .
\eqn{D_T}
With the help of this covariant derivative one easily finds
the following identities:
\eqa
\frac{dF^{A}(t)}{dt}\=dA^{A}+tf^{ABC}A^{B}A^{C} \ , \non
D_{t}A^{A}\=dA^{A}+f^{ABC}A^{B}(t)A^{C} \non
\=dA^{A}+tf^{ABC}A^{B}A^{C}=\frac{dF^{A}(t)}{dt} \ ,
\eqan{IDENT_1}
and the Bianchi identity
\eqa
D_{t}F^{A}(t)\=dF^{A}(t)+f^{ABC}A^{B}(t)F^{C}=0 \ .
\eqan{IDENT_2}
We are now able to derive the Chern-Simons formula
by starting with the corresponding Chern
character in dimension $2k$ and using above identities:
\eqa
Tr(F^{k})\=Tr\int\limits_{0}^{1} dt \frac{d}{dt} F(t)^{k} \non
\=k~Tr\int\limits_{0}^{1} dt \lp\frac{dF(t)}{dt}\rp F(t)^{k-1} \non
\=k~Tr\int\limits_{0}^{1} dt D_{t}\lp AF(t)^{k-1}\rp \non
\=kd~Tr\int\limits_{0}^{1} dt~AF(t)^{k-1} \non
\=d\O^{0}_{2k-1} \ ,
\eqan{CHERN}
where the Chern-Simons term in $(2k-1)$ dimensions is now given by
\eq
\O^{0}_{2k-1}=k~Tr\int\limits_{0}^{1} dt~AF(t)^{k-1} \ .
\eqn{CHERN_SIMONS}

\section*{B~~~0-form cocycles for Chern-Simons terms
and for gauge anomalies}

\setcounter{equation}{0}
\renewcommand{\theequation}{B.\arabic{equation}}

In this appendix we derive first the general
0-form cocycle $\O^{2k-1}_{0}$ \equ{CS_FORMULA_ZERO}
in $(2k-1)$ dimensions, which leads by applying the operator $\d$
to the Chern-Simons formula \equ{CHERN_SIMONS}.
In a second step we show that
this cocycle is BRST-invariant and non-trivial, i.e. that
$\O^{2k-1}_{0}$ identifies a cohomology class of the BRST operator.

Since the decomposition $\d$ \equ{DECETA} acts exclusively on the
local translation ghost $\h^{a}$ one has only to substitute the
forms in \equ{A_T} and \equ{F_T} by the corresponding ghost fields.
Taking into account \equ{C_HAT} and \equ{DELTA_C_HAT}, we make
the following substitutions:
\eqa
&&A^{A} \longrightarrow -\hat{c}^{A} \ , \non
&&F^{A} \longrightarrow \hat{F}^{A} \ .
\eqan{SUBST}
The associated interpolated ghost field strength $\hat{F}^{A}(t)$
is given by
\eqa
\hat{F}^{A}(t)\=t\hat{F}^{A}
+\frac{1}{2}t(t-1)f^{ABC}\hat{c}^{B}\hat{c}^{C} \ .
\eqan{F_HAT_T_APPENDIX}
One can now conclude from \equ{CHERN_SIMONS} to the 0-form
cocycle $\O^{2k-1}_{0}$ in $(2k-1)$ dimensions according to
\eqa
\O^{2k-1}_{0}\=-k~Tr\int\limits_{0}^{1} dt~\hat{c}
\hat{F}(t)^{k-1} \ .
\eqan{CS_FORMULA_ZERO_APPENDIX}
With the help of eq.\equ{HICO} one easily regains the corresponding
$(2k-1)$-dimensional Chern-Simons term:
\eqa
\O^{0}_{2k-1}\=\frac{\d^{2k-1}}{(2k-1)!}\O^{2k-1}_{0} \non
\=k~Tr\int\limits_{0}^{1} dt~AF(t)^{k-1} \ .
\eqan{CS_GENERAL_APPENDIX}

To prove the BRST invariance of \equ{CS_FORMULA_ZERO_APPENDIX}
we recall the BRST transformations of $\hat{c}^{A}$ and
$\hat{F}^{A}$:
\eqa
s\hat{c}^{A}\=\frac{1}{2}f^{ABC}\hat{c}^{B}\hat{c}^{C}
-\hat{F}^{A} \ , \non
s\hat{F}^{A}\=f^{ABC}\hat{c}^{B}\hat{F}^{C} \ ,
\eqan{BRST_GHOSTS_COMP}
or in matrix notation
\eqa
s\hat{c}\=\hat{c}\hat{c}-\hat{F} \ , \non
s\hat{F}\=\lc\hat{c},\hat{F}\rc \ ,
\eqan{BRST_GHOSTS}
with $\hat{c}=\hat{c}^{A}T^{A}$ and $\hat{F}=\hat{F}^{A}T^{A}$,
whereby $T^{A}$ are the generators of the gauge group.
Moreover, eq.\equ{F_HAT_T_APPENDIX} can be expressed
in a way as below:
\eq
\hat{F}(t)=t^{2}\hat{c}\hat{c}-ts\hat{c} \ .
\eqn{F_HAT_T_RE-EXPRESS}
Analogous to \equ{D_T} we define an interpolating
covariant BRST operator $S_{t}$ according to
\eq
S_{t}=s-\hat{c}(t)~~~,~~~\hat{c}(t)=t\hat{c} \ ,
\eqn{S_T}
which obeys the following identities:
\eqa
&&S_{t}\hat{F}(t)=0 \ , \non
&&S_{t}\hat{c}=s\hat{c}-2t\hat{c}\hat{c}
=-\frac{d}{dt}\hat{F}(t) \ .
\eqan{ID_APPENDIX}
Finally, by using above relations one gets:
\eqa
s\O^{2k-1}_{0}\=-k~Tr\int\limits_{0}^{1} dt~s\lp\hat{c}
\hat{F}(t)^{k-1}\rp \non
\=-k~Tr\int\limits_{0}^{1} dt~\lc S_{t}\lp\hat{c}
\hat{F}(t)^{k-1}\rp+\lac\hat{c}(t),\hat{c}
\hat{F}(t)^{k-1}\rac\rc \non
\=k~Tr\int\limits_{0}^{1} dt~\lc \frac{d\hat{F}(t)}{dt}
\hat{F}(t)^{k-1}\rc \ .
\eqan{BRST_INVARIANCE_ZERO}
We remark that the second term in the second line of
\equ{BRST_INVARIANCE_ZERO} is identically equal to zero.
The BRST transformation of $\O^{2k-1}_{0}$ is now given by
\eqa
s\O^{2k-1}_{0}\=\int\limits_{0}^{1} dt~\frac{d}{dt}
Tr\lp\hat{F}^{k}(t)\rp \non
\=\left. Tr\lp\hat{F}^{k}(t)\rp\right|^{1}_{0} \non
\=Tr \hat{F}^{k} \ .
\eqan{BRST_ZERO}
{}From the definition \equ{C_HAT} and taking into account
that in an $N$-dimensional spacetime the product of $(N+1)$
ghost fields $\h$ automatically vanishes, it follows that the
0-form cocycle $\O^{2k-1}_{0}$ given by
\equ{CS_FORMULA_ZERO_APPENDIX}
is BRST-invariant:
\eqa
s\O^{2k-1}_{0}\=\frac{1}{2^{k}}Tr\lp F_{m_1 n_1}F_{m_2 n_2}
.....F_{m_k n_k}\h^{m_1}\h^{n_1}\h^{m_2}\h^{n_2}
.....\h^{m_k}\h^{n_k} \rp =0 \ .
\eqan{BRST_INV}

To finish the proof that $\O^{2k-1}_{0}$ belongs to a cohomology
class of the BRST operator one has also to show that it is a
non-trivial solution, i.e.
\eq
s\O^{2k-1}_{0}=0~~~,~~~\O^{2k-1}_{0}\not= s\hat{\O}^{2k-2}_{0} \ .
\eqn{NONTRIVIAL_SOLUTION}
For this reason we define a filtering operator $\NN$ as follows:
\eq
\NN=\th\frac{\6}{\6\th}+c\frac{\6}{\6 c}+\o\frac{\6}{\6\o}
+T\frac{\6}{\6 T}+R\frac{\6}{\6 R}+F\frac{\6}{\6 F} \ ,
\eqn{FILTER_OPERATOR}
which induces a separation on the BRST transformations
\equ{0_FORMS} and \equ{FORMZERO}
of the elementary 0-form fields according to
\eqa
s^{(0)}\h^{a}\=0 \ , \non
s^{(0)}\th^{a}_{~b}\=-\h^{k}\6_{k}\th^{a}_{~b} \ , \non
s^{(0)}c^{A}\=-\h^{k}\6_{k}c^{A} \ , \non
s^{(0)}\o^{a}_{~bm}\=-\6_{m}\th^{a}_{~b}
-\h^{k}\6_{k}\o^{a}_{~bm} \ , \non
s^{(0)}A^{A}_{m}\=-\h^{k}\6_{k}A^{A}_{m} \ , \non
s^{(0)}T^{a}_{mn}\=-\h^{k}\6_{k}T^{a}_{mn} \ , \non
s^{(0)}R^{a}_{~bmn}\=-\h^{k}\6_{k}R^{a}_{~bmn} \ , \non
s^{(0)}F^{A}_{mn}\=-\h^{k}\6_{k}F^{A}_{mn} \ .
\eqan{SEPARATION_BRST}
The general 0-form cocycle $\O^{2k-1}_{0}$ can be rewritten as
\eqa
\O^{2k-1}_{0}\=-k~Tr\int\limits_{0}^{1} dt~\lac\hat{c}\lp
t\hat{F}+t(t-1)\hat{c}\hat{c}\rp^{k-1}\rac \non
\=-k~Tr\int\limits_{0}^{1} dt~t^{k-1}\lac(t-1)^{k-1}\hat{c}^{2k-1}
+~.....\rac \non
\=-k~\int\limits_{0}^{1} dt~\D(t) \ .
\eqan{GEN_REW}
The filtration \equ{FILTER_OPERATOR} induces now a separation
on $\D(t)$
\eqa
\D(t)\=\D^{(0)}(t)+\sum_{p \ge 1}^{\bar{p}} \D^{(p)}(t) \ ,
\eqan{SEP_FILTER}
where $\D^{(0)}(t)$ is given by
\eqa
\D^{(0)}(t)\=\a(t)\D^{(0)}=\a(t)Tr\lp A_{m}\h^{m}\rp^{2k-1} \ .
\eqan{DELTA_NULL}
A possible triviality of $\O^{2k-1}_{0}$ would
imply at the level of the integrands
\eq
\D^{(0)}=s^{(0)}\hat{\D}^{(0)}
=s^{(0)}Tr\lp A_{m}\h^{m}\rp^{2k-2} \ ,
\eqn{INT_LEVEL}
but from the fith eq. of \equ{SEPARATION_BRST} follows that
above equation has no solution. Therefore the 0-form cocycle
$\O^{2k-1}_{0}$ is a non-trivial solution of the BRST cohomology
\eq
\O^{2k-1}_{0}\not= s\hat{\O}^{2k-2}_{0} \ ,
\eqn{NONTRIVIAL_RESULT}
because of the fact that the cohomology of $s$ is isomorphic to
a subspace of the cohomology of $s^{(0)}$~\cite{dixon,bandel}.


\end{document}